\RequirePackage{lineno}
\documentclass[a4paper,11pt]{article}
\pdfoutput=1 

\usepackage{jheppub} 

\usepackage[T1]{fontenc} 

\usepackage{xspace}
\usepackage{graphicx}
\usepackage{caption}
\usepackage{subcaption}
\usepackage{relsize}

\input{defs.tex}

\newcommand{\comment}[1]{}

\newcommand{\equref}[1]       {Eq.~(\ref{eq:#1})}
\newcommand{\Equref}[1]       {Eq.~(\ref{eq:#1})}

\newcommand{\secref}[1]       {Section~\ref{sec:#1}}

\title{\boldmath Particle-level pileup subtraction for jets and jet shapes}

\author[a]{Peter Berta,}
\author[a]{Martin Spousta,}
\author[b]{David W. Miller,}	
\author[a]{Rupert Leitner}

\affiliation[a]{Faculty of Mathematics and Physics, Charles University in Prague,\\V Hole\v sovi\v ck\' ach 2, Prague, Czech Republic}
\affiliation[b]{Enrico Fermi Institute, University of Chicago, Chicago, IL, USA}

\emailAdd{berta@ipnp.troja.mff.cuni.cz}
\emailAdd{Martin.Spousta@cern.ch}
\emailAdd{David.W.Miller@uchicago.edu}
\emailAdd{Rupert.Leitner@mff.cuni.cz}

\abstract{We present an extension to the jet area-based \pileup subtraction for both jet kinematics and jet shapes. A particle-level approach is explored whereby the jet constituents are corrected or removed using an extension of the methods currently being employed by the LHC experiments. Several jet shapes and nominal jet radii are used to assess the performance in simulated events with \pileup levels equivalent to approximately 30 and 100 interactions per bunch crossing, which are characteristic of both the LHC Run I and Run II conditions. An improved performance in removing the \pileup contributions is found when using the new subtraction method. The performance of the new procedure is also compared to other existing methods.}


\begin{document} 

\maketitle
\flushbottom

\section{Introduction}
\label{sec:intro}

The intense environment created by the high luminosity of the Large Hadron Collider (LHC) necessitates novel methods for isolating, mitigating, and, where possible, correcting for the contributions of multiple uncorrelated proton-proton interactions (\pileup) to the measured hadronic final state. \Pileup has a substantial impact on both jet kinematics and substructure, thereby degrading critical tools for identifying new physics via highly boosted hadronic decays of $W$, $Z$, and Higgs bosons, or top quarks.

The last few years has seen the development of several effective tools for \pileup mitigation and removal. Simple subtraction techniques remove a constant offset from the measured transverse momentum that is proportional to the number of observed \pileup events~\cite{babt:ATLAS}. So-called \textit{grooming techniques} such as filtering~\cite{aakh:Butterworth}, pruning~\cite{aaki:Ellis}, and trimming~\cite{aakj:Krohn}, actively remove potential \pileup constituents from jets. Other approaches, such as the \textit{jet cleansing method}~\cite{aakg:Krohn} or \textit{charged hadron subtraction}~\cite{aakl:CMS}, use tracking information to identify a given hadronic energy deposition with charged particles originating in \pileup interactions.

Techniques that utilize event-by-event and jet-by-jet information to determine the extent of contamination from \pileup provide a new approach to perform jet physics at very high luminosities. The \textit{area-based subtraction} procedure~\cite{aabc:Cacciari} corrects the jet \fourmomentum and it is extended to account for hadron masses in Ref.~\cite{aaig:Soyez}. The \textit{shape-expansion method}~\cite{aaig:Soyez} provides general approach to correct jet shapes.


The extension of the area-based subtraction procedure that we propose here allows for a particle-by-particle approach to this concept. We find improved performance in removing the contributions due to \pileup using this new procedure even for previously intractable jet shape observables (such as planar flow). Furthermore, this approach provides the possibility to perform \pileup subtraction without explicit consideration of a specific jet algorithm, reducing the constraints and increasing the flexibility of the jet area-based subtraction procedure overall. Therefore, this method may be used also in heavy ion physics where the jet reconstruction is challenging due to sizable underlying event \cite{aadu:Apolinario, aaih:Spousta}.

\section{Subtraction algorithm}
\label{sec:algo}
  The novel feature of the approach described here is the \textit{local} subtraction of \pileup at the level of individual jet constituents.
  In contrast to the area-based subtraction and the shape-expansion method, the constituent-level subtraction is performed particle-by-particle, thereby correcting both the \fourmomentum of the jet and its substructure, simultaneously. This is achieved by combining the kinematics of particles within a specific jet with the kinematics of soft ``negative'' particles that are added to balance the \pileup contribution.



  The basic ingredient of the particle-level subtraction is the \pileup energy density estimation which is identical to that used in the shape-expansion method proposed in Ref.~\cite{aaig:Soyez}. 
  The contamination due to \pileup is described in terms of the transverse momentum density $\rho$ and mass density $\rhom$. The expected \pileup deposition in a small region of $\Dy\Dphi$ is expressed by the \fourmomentum
  %
\begin{equation}
  p^{\mu}_{\rm \pileup} = [\rho \cos{\phi},\rho  \sin{\phi},(\rho + \rhom)  \sinh{y}, (\rho + \rhom) \cosh{y}] \cdot \Dy\Dphi,
  \label{eq:pileup} 
\end{equation} 
  %
  where the \pileup \pt and mass densities, $\rho$ and \rhom, are assumed to be weakly dependent on rapidity $y$ and azimuth $\phi$. In the shape-expansion method, all particles (or effective particles such as calorimeter towers \cite{aaio:Bellettini} or clusters \cite{aaip:Lampl}) in the event are grouped into \textit{patches} in order to estimate the densities used in \equref{pileup}. The patches are defined by jets reconstructed using the \kt algorithm \cite{aaiw:Ellis, aaiy:Catani}.
  The transverse momentum, $\ptpatch$, and mass, $\mdeltapatch$, of each patch is determined by summing over all particles within that patch:
  %
\begin{equation}
  \ptpatch = \sum_{i\in\mathrm{patch}}\pti, ~~~~~
  \mdeltapatch = \sum_{i\in\mathrm{patch}}\left(\sqrt{m_i^2+\pti^2} - \pti\right),
  \label{eq:patches}
\end{equation}
  \noindent where \pti and $m_i$ are the transverse momentum and mass of particle $i$, respectively. 
  Each patch covers certain area $\areapatch$ in the ($y-\phi$) plane. The overall background \pt and mass densities are estimated as
  %
\begin{equation}
  \rho = \mathrm{median}_{\mathrm{patches}}\left\{ \frac{\ptpatch}{\areapatch}\right \}, ~~~~~\rhom = \mathrm{median}_{\mathrm{patches}}\left\{ \frac{\mdeltapatch}{\areapatch}\right \},
  \label{eq:rho}
\end{equation}
  %
  although several modifications exist, including $y$-dependent $\rho$ and \rhom \cite{aain:Cacciari}. 
  The estimation of the background densities is the first step which needs to be followed by a scheme by which to subtract a specified amount of those densities.



  

In our approach, massless particles with very low momentum are incorporated into the event such that they uniformly cover the \ensuremath{y - \phi} plane with high density. These soft particles are referred to as \textit{ghosts} and they are most commonly used to define the area of a jet \cite{aabb:Cacciari} or to perform the shape-expansion correction. Each ghost covers a certain fixed area, \Aghost, in the \ensuremath{y - \phi} plane which is defined by the ghost number density ($\Aghost$ is its inverse). The \fourmomentum of each particle or ghost is expressed as
  %
\begin{equation}
  p^{\mu} = [\pt \cos{\phi},\pt  \sin{\phi},(\pt + \mdelta)  \sinh{y}, (\pt + \mdelta) \cosh{y}],
  \label{eq:particle}
\end{equation}
  %
  where $\mdelta = \sqrt{m^2 + \pt^2} - \pt$ (in what follows, we will use superscript $g$ to denote the kinematic variables of ghosts). After adding ghosts into the event, the jet clustering algorithm runs over all particles and ghosts delivering the same jets as in the case without the ghosts. Now, the jets contain except the real particles also ghosts which can be used to correct for the \pileup contribution within each jet.
  \Equref{pileup} is translated into the \fourmomentum of each ghost by identifying the transverse momentum \ghostpt\ and mass \ghostmdelta\ with the amount of \pileup within area \Aghost:
  %
\begin{equation}
\begin{split}
  & \ghostpt     = \Aghost\cdot\rho, \\
  & \ghostmdelta = \Aghost\cdot\rhom.
  \label{eq:new_ghosts}
\end{split}
\end{equation}
  %
An iterative procedure is used to define the scheme for calculating the specified amount of transverse momentum and mass \mdelta to subtract from each jet constituent. For each pair of particle $i$ and ghost $k$, a matching scheme is implemented using the distance measure, $\DeltaRik$, defined as
  %
\begin{equation}
  \DeltaRik=\pti^\alpha \cdot \sqrt{\left(\yi-\ghostyk\right)^2+\left(\phii-\ghostphik\right)^2}.
  \label{eq:deltaRik}
\end{equation}
  %
  For complete generality, $\alpha$ is allowed to be any real number, but is taken to be zero in the studies performed here. The list of all distance measures, $\{\DeltaRik\}$, is sorted from the lowest to the highest values. The \pileup removal proceeds iteratively, starting from the particle-ghost pair with the lowest $\DeltaRik$. At each step, the momentum \pt and mass \mdelta of each particle $i$ and ghost $k$ are modified as follows.
  %
\begin{equation}
\begin{split}
\text{If}~ \pti\geq\ghostptk: ~~~~~
   &\pti\longrightarrow\pti-\ghostptk, \\
   & \ghostptk\longrightarrow 0; \\
\text{otherwise:}~~~~~~
   & \pti\longrightarrow 0, \\
   & \ghostptk\longrightarrow \ghostptk-\pti. \\
\end{split}
\quad \raisebox{-2ex}{\scalebox{2.5}{\Bigg\vert}} \quad
\begin{split}
\text{If}~ \mdeltai\geq\ghostmdeltak: ~~~~~
   &\mdeltai\longrightarrow\mdeltai-\ghostmdeltak, \\
   & \ghostmdeltak\longrightarrow 0; \\
\text{otherwise:}~~~~~~
   & \mdeltai\longrightarrow 0, \\
   & \ghostmdeltak\longrightarrow \ghostmdeltak-\mdeltai. \\
\end{split}
  \label{eq:correction}
\end{equation}
  %
The azimuth and rapidity of the particles and ghosts remain unchanged. The iterative process is terminated when the end of the sorted list is reached. Alternatively, a threshold $\DeltaRmax$ can be introduced to stop the iterations when $\DeltaRik > \DeltaRmax$. In principle, introducing the $\DeltaRmax$ threshold also guarantees that only ghosts neighbouring a given particle are used to correct the kinematics of that particle. Particles with zero transverse momentum after the iterative process are discarded and the \fourmomentum of a given jet is recalculated following a desired recombination scheme (commonly, the \fourmomentum recombination scheme is used \cite{aaiq:Blazey}). It can happen that after the subtraction no real particle remains. This may be a signal that such a jet originates from \pileup.

  The scalar subtraction in \equref{correction} is chosen instead of 4-momentum subtraction since it allows to reduce the local differences between the actual background deposit and its estimate from \equref{pileup}\footnote{An alternative form of \equref{correction} with subtracting the \fourmomenta would lead to a corrected jet with the same \fourmomentum as if the area-based subtraction from Ref.~\cite{aaig:Soyez} was applied which follows from the additiveness of \fourmomenta. The performance of such particle-level correction will be explored in an upcoming study.}. The scalar subtraction also eliminates the occurrence of unphysical negative squared masses which may be present when subtracting 4-momenta. Furthermore, the scalar summation is also used in the calculation of the transverse momentum of patches in \equref{patches} and thereby it avoids the dependence of background densities in \equref{rho} on the shape and size of the patches. 

  The above described subtraction procedure -- referred to as \textit{\subtraction} -- corrects both the \fourmomentum of a jet as well as its substructure. The \subtraction works equally well when applied directly to Monte Carlo truth particles from simulation as when applied to a coarse pseudo-detector grid over which the energy from the truth particles is distributed (see Sec.~\ref{sec:shapes}). An important feature of the algorithm is that it preserves longitudinal invariance -- an arbitrary jet after the correction and a subsequent boost in the direction of colliding beams has the same constituents as the same jet which is first boosted and then corrected. The algorithm can be easily extended to account for rapidity dependence of background densities $\rho$ and \rhom in \equref{new_ghosts}.

It is also straightforward to extend the method further to the whole event instead of correcting just particles within a jet. Jet finding can then be performed using the subtracted event. Global event shapes can also benefit from the correction in addition to individual jet observables. The performance and resolution of missing transverse energy calculated from calorimeter energy deposits may also improve, thereby enhancing several searches for physics beyond the Standard Model. Studies of whole-event constituent subtraction are left to future work, as well as testing directly within the experimental communities. 

An important advantage of the \subtraction is the speed -- it can be as much as twenty times faster compared to the shape-expansion method, depending on the type of the jet shape and the nominal jet radius. Furthermore, the shape-expansion correction must be determined for each jet shape in consideration, whereas the \subtraction approach provides a corrected set of constituents, from which any shape may be determined. Corrected constituents may also then be used as inputs to jet grooming and tagging algorithms, e.g. the top-quark tagging using the shower deconstruction method~\cite{aakd:Soper}. In comparison to the jet cleansing method or charged hadron subtraction used by the CMS experiment, the constituent subtraction does not require any knowledge about the connection of each charged particle with the signal vertex or \pileup vertices, though such a knowledge might in principle be used to further enhance capabilities of the algorithm.

The \subtraction procedure has following free parameters: $\Aghost$, $\DeltaRmax$, and $\alpha$. The basic recommended settings are: $\Aghost=0.01$, $\DeltaRmax \rightarrow \infty$, and $\alpha=0$. These settings were used in the performance studies presented in Sec.~\ref{sec:performance}. The subtraction is stable with respect to varying $\Aghost$. The variation of $\Aghost$ by a factor of two does not lead to a change in any of the studied quantities that would be significant with respect to the statistical uncertainty shown on plots in Sec.~\ref{sec:performance}. Introducing a finite $\DeltaRmax$ may improve the performance of the correction and the speed of the algorithm when running over the full event. The configuration with $\alpha > 0$ prefers to subtract the lower \pt constituents which more often originate from background. This configuration will not be discussed in this paper but it appears to lead to an improvement in the correction of some of the jet shapes.

The software for the \subtraction will be implemented as a part of the \texttt{FastJet Contrib} project \cite{aaiv:noauthor}.

%
%
%

\section{Performance of the subtraction}
\label{sec:performance}
The \subtraction algorithm corrects both the jet kinematics (\pt\ and mass) and the jet shapes. Studies of the algorithm performance for \pt are discussed in \secref{kinematics} along with comparisons to the area-based subtraction which follows Ref.~\cite{aaig:Soyez} where the \pileup \fourmomentum (\ref{eq:pileup}) is subtracted using the jet area \fourvector $A^{\mu}$ as
\begin{equation}
  p^{\mu}_{\mathrm{corr}} = [p^x - \rho A^x,p^y - \rho A^y,p^z - (\rho + \rhom)  A^y, E - (\rho + \rhom)  A^E].
\label{eq:correctedFourMomentum} 
\end{equation}
  The performance of the subtraction applied to both the jet mass and several jet shapes is presented in \secref{shapes}. Comparisons of the \subtraction approach with the shape-expansion method~\cite{aaig:Soyez} are also presented.

The studies presented are performed using events generated with PYTHIA 8.180, tune 4C~\cite{aaat:Sjostrand, aaiz:Sjostrand} but without any detector simulation. The effect of additional proton-proton collisions is emulated by using inclusive events (often referred to as ``minimum bias'' events) overlaid with the hard scattering interaction, which are also generated with PYTHIA 8.180. The CTEQ 5L, LO parton density functions~\cite{aaka:Lai}, configured to simulate the LHC conditions at $\sqrt{s} = 8$~TeV, are used for all event generation. Two processes are simulated without underlying event: di-jet events covering the \pt\ range of 10-800 GeV and events with boosted top quarks from decay $Z' \rightarrow t\bar{t}$ of hypothetical boson $Z'$ with mass of $1.5\TeV$. The performance of the subtraction is tested using jets clustered with the \akt algorithm~\cite{aaba:Cacciari} with the distance parameter $R=0.7$ or $R=1.0$ and jets clustered with Cambridge-Aachen (\CamKt) algorithm~\cite{aair:Dokshitzer} with $R=1.2$. These are representative jet definitions for both the ATLAS~\cite{babu:ATLAS} and CMS~\cite{aakm:CMS} experimental collaborations. The number of \pileup events, \npu, has a Poisson distribution with a mean \avgnpu. Two \pileup conditions are simulated, $\avgnpu=30$ and $\avgnpu=100$. These \pileup configurations represent realistic conditions for LHC Run I and upcoming LHC Run II as well as for the high luminosity LHC running \cite{aaim:Bruning}. On average, the \pileup contribution to the hard-scatter event can be destribed through mean value of transverse momentum densities, $\avgrho$, and \pileup fluctuations characterised by standard deviation, $\sigmarho$. For the used configuration $\avgnpu=100$, these quantities are $\avgrho\approx75\GeV$ and $\sigmarho\approx13\GeV$.

  

  All jet finding and background estimation is performed using \texttt{FastJet} 3.0.6~\cite{aaiu:Cacciari, aain:Cacciari}. The shape-expansion correction is performed using \texttt{FastJet Contrib} 1.003~\cite{aaiv:noauthor}. The patches in \equref{rho} are obtained by clustering particles with the \kt algorithm~\cite{aaiw:Ellis, aaiy:Catani} with distance parameter $R=0.4$. The non-negligible dependence of the background densities $\rho$ and $\rhom$ on rapidity impacts each of the corrections methods discussed below. Consequently, in order to focus the comparisons and performance evaluations, only patches with rapidity $|y|<2.0$ are used in \equref{rho} and jets are required to be fairly central, with $|\eta|<2.0$.



\subsection{Jet kinematics}
\label{sec:kinematics}

The ability of the subtraction to correctly recover the kinematics of the jet can be characterized in terms of following quantities: jet momentum response, jet momentum resolution, jet position resolution, and jet finding efficiency. These quantities are commonly used to evaluate the performance of the jet reconstruction, see e.g. Refs.~\cite{babr:ATLAS, aaij:CMS, aaik:CDF, aail:D0}. In this section we evaluate these quantities both for the \subtraction and area-based subtraction.

\begin{figure}[tp]
\centering 
\begin{center}
\includegraphics[width=0.328\textwidth]{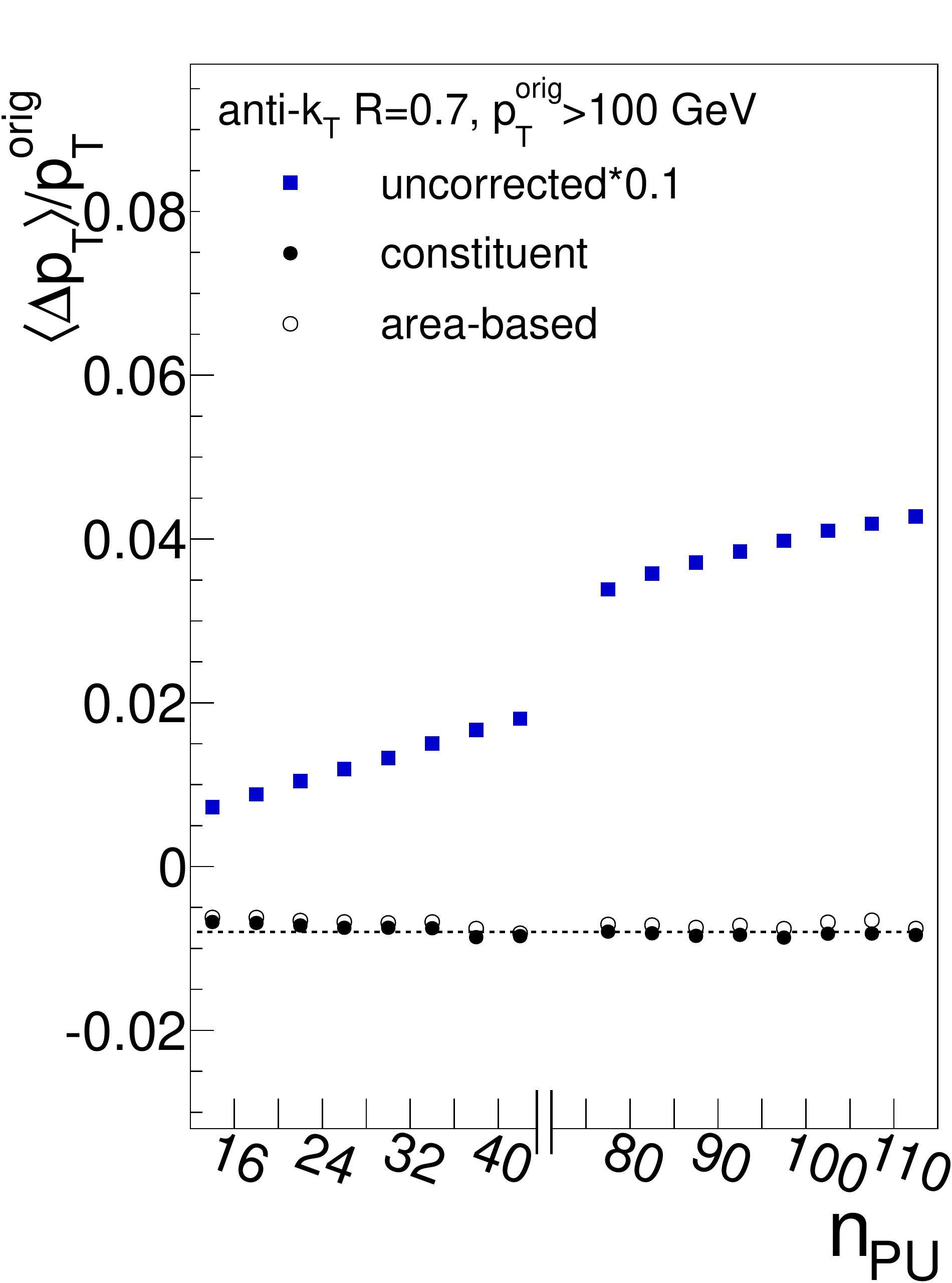}
\includegraphics[width=0.328\textwidth]{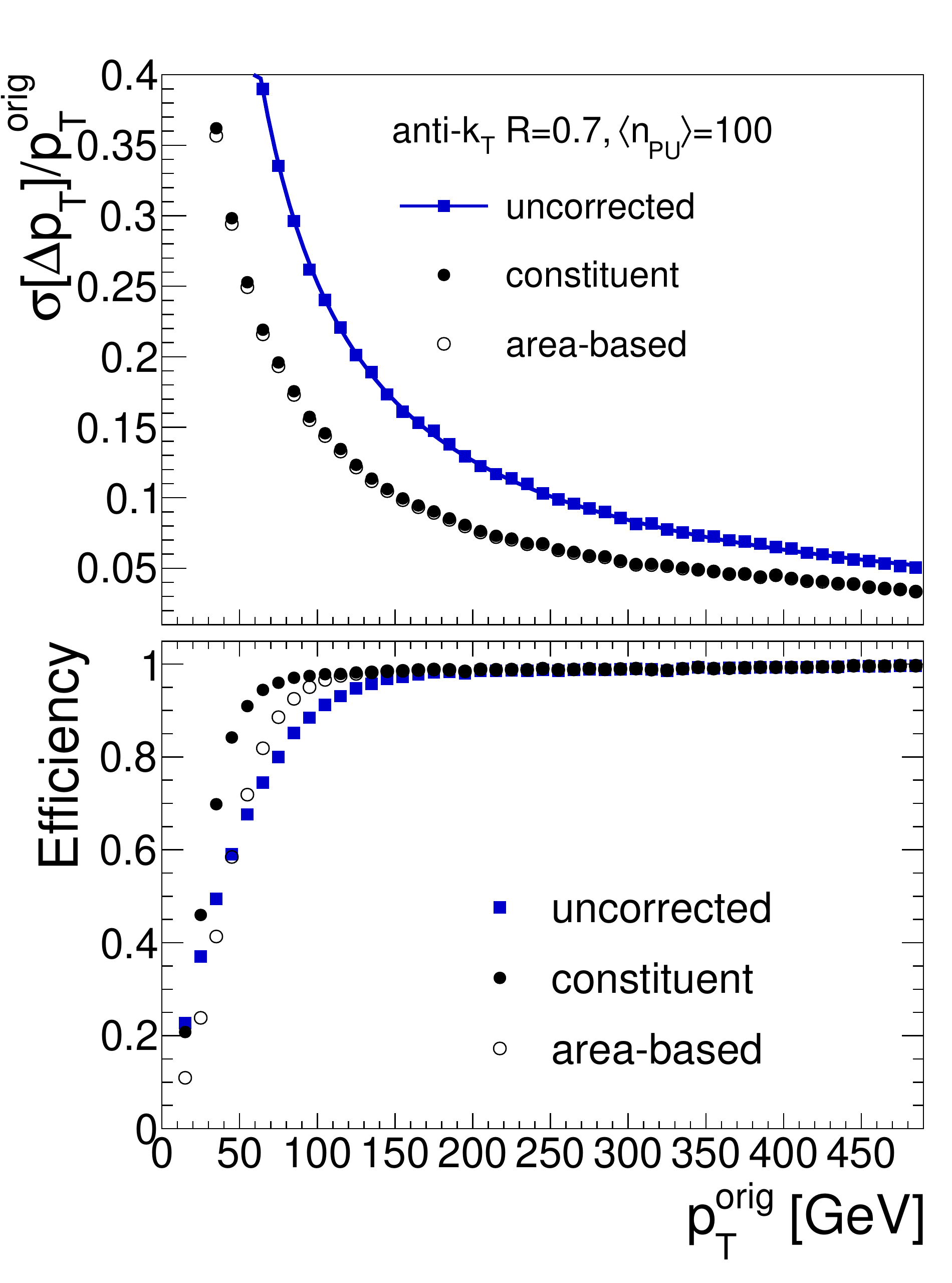}
\includegraphics[width=0.328\textwidth]{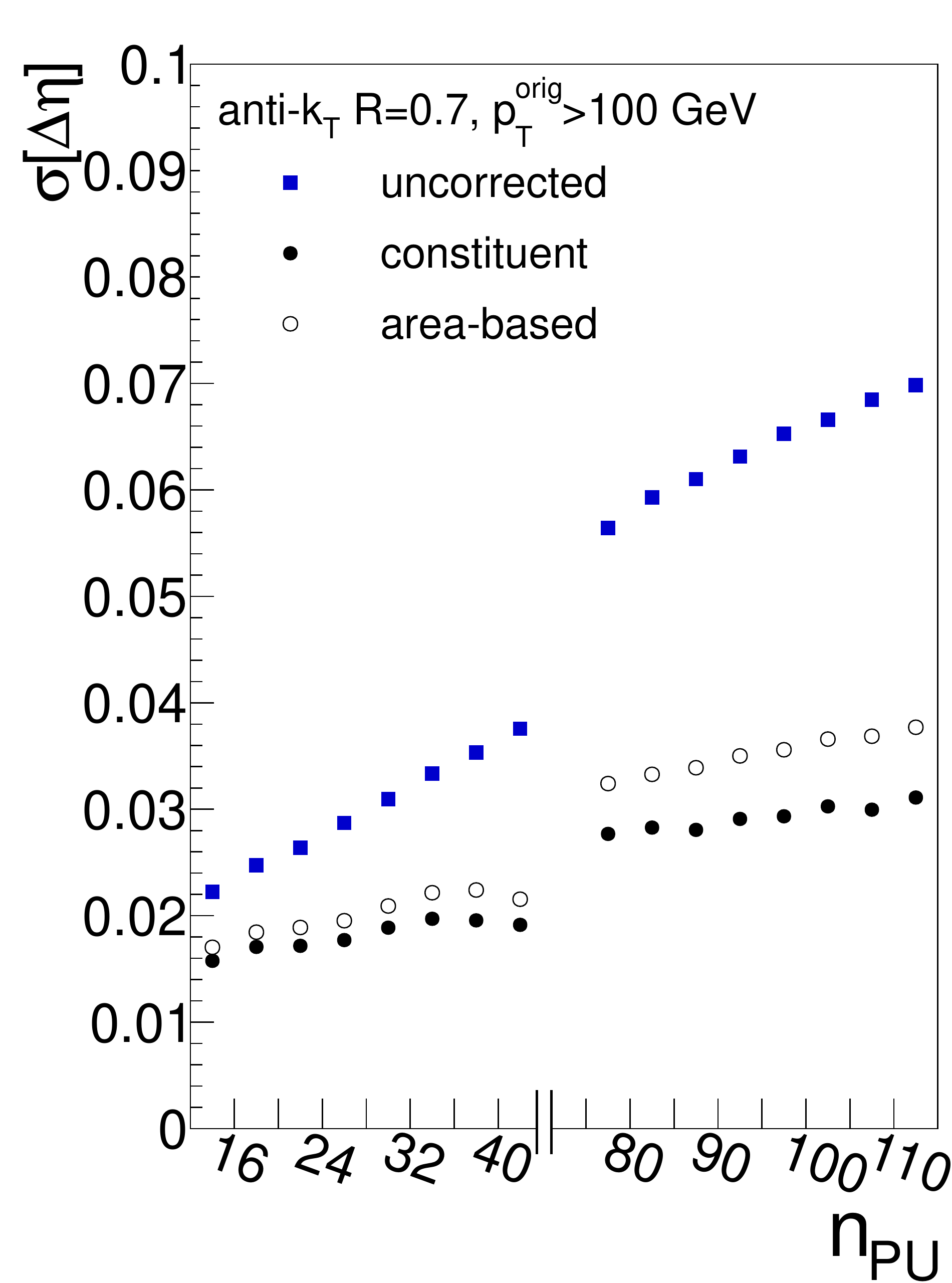}
\end{center}
  \caption{\label{fig:kinematics} 
  Jet momentum response (left). Jet momentum resolution (upper middle) and jet finding efficiency (lower middle). Jet position resolution (right). Jets prior the subtraction denoted as \pileup (square markers) are compared to jets corrected by the \subtraction (full circles) and jets corrected by the area-based subtraction (open circles). The dashed line in the left panel shows a constant at $-0.8\%$ to guide the eye. Curve in the upper middle panel represents a fit of the jet momentum resolution by $c/\pt$ resulting $c\approx25\GeV$.}
  \end{figure}

 The first quantity characterizing the basic performance of the subtraction is the jet momentum response. It can be defined as ${\langle \Delta \pt \rangle}/{\ptorig} = {\langle \pt - \ptorig \rangle}/{\ptorig}$ where \ptorig\ is the original jet momentum with no \pileup and \pt\ is either the \pileup corrected jet momentum or the momentum of uncorrected jet that is jet clustered in the presence of \pileup with no subtraction. This quantity is also often referred to as the jet energy scale. In the optimal situation, the jet momentum response should be zero which means that, on average, the algorithm can reconstruct the same \pt\ as with no \pileup. Left panel of Fig.~\ref{fig:kinematics} shows the jet momentum response as a function of the number of \pileup collisions, \npu, that is the size of the \pileup. The jet momentum response of subtracted jets differs from zero by less then a 1\%. 
  The jet momentum response of subtracted jets is stable with respect to the \pileup which is a crucial condition for the jet reconstruction. Without satisfying this condition any cut applied on the jet \pt\ would lead to a choice of different subset of jets depending on the size of the \pileup. Small deviation from zero of the jet momentum response is resulting from ignoring the rapidity dependence of the \pileup density $\rho$ and $\rhom$ and from a small average bias of \pileup densities by a presence of the hard-scatter event. Such small deviation can be easily corrected after the jet reconstruction by multiplying the jet momentum by a correction factor. The \subtraction performs equally well as the area-based subtraction.

  The small deviation of jet momentum response from zero for corrected jets can be contrasted with the jet momentum response of uncorrected jets -- in the case of low \pileup scenario the momentum response is 10-20\%, in the case of high \pileup scenario the jet momentum response is around 40\%. This means that in the high \pileup case, there is on average approximately 40 GeV of the \pileup background underneath each jet leading to a reconstruction of a typical 100 GeV jet as a 140 GeV jet if performing no subtraction.


The second quantity characterizing the basic performance of the subtraction is the jet momentum resolution defined as ${\sigma[ \Delta \pt ]}/{\ptorig} = {\sigma[ \pt - \ptorig ]}/{\ptorig}$ where $\sigma[...]$ denotes the standard deviation. The jet momentum resolution is dictated by the presence of fluctuations in the underlying \pileup background leading to dependence ${\sigma[ \Delta \pt ]}/{\ptorig} = c/\ptorig$ where $c$ is a constant. The jet momentum resolution is shown in the upper middle panel of Fig.\ref{fig:kinematics}. The fit of the jet momentum resolution of uncorrected jets by $c/\ptorig$ leads to $c \approx 25\GeV$ which results from the magnitude of \sigmarho and \pileup fluctuations in $\eta-\phi$ plane within each particular event. The \subtraction and area-based subtraction have similar jet momentum resolution while both methods significantly improve it.



  The jet finding efficiency is defined as the number of original jets having a matching corrected (or uncorrected) jet divided by the number of original jets. The matching criterion is the distance in the $\eta-\phi$ plane between the original jet and corrected (or uncorrected) jet satisfying the condition $\Delta R = \sqrt{ \Delta\eta^2 + \Delta\phi^2} < 0.2$. This quantity is plotted in the lower middle panel of Fig.~\ref{fig:kinematics}. It shows that in the case of high \pileup events it is difficult in principle to reconstruct the jets with $\ptorig < 50$~GeV due to the presence of sizable fluctuating background. The jet efficiency is better for the \subtraction than for the area-based subtraction.

  The last quantity evaluated is the jet position resolution which characterizes the ability to recover the original jet axis in pseudorapidity, $\eta$, or azimuth, $\phi$. The jet position resolution in $\eta$ and $\phi$ are similar. The jet position resolution in $\eta$, $\sigma[\Delta\eta]$, is defined as the standard deviation of the difference between the original jet $\eta$ position and $\eta$ position of the corrected (or uncorrected) jet. The $\sigma[\Delta\eta]$ is plotted in the right panel of Fig.~\ref{fig:kinematics} where a clear difference between jets with \pileup contribution and corrected jets is present. Jets corrected by the area-based subtraction have slightly worse jet position resolution than jets corrected by the \subtraction.

Based on the analysis of the basic performance, we can conclude the \subtraction has a good ability to correct for the \pileup background and to recover the original jet kinematics even in the presence of a sizable \pileup. The \subtraction has generally similar performance as the area-based subtraction. A slightly better performance in terms of jet position resolution and jet efficiency may be attributed to the correction of the jet internal structure which is done by the \subtraction. The ability of the \subtraction to correct the jet internal structure is discussed in \secref{shapes}.

\subsection{Jet shape definitions}
\label{sec:definitions}
The ability of the \subtraction method to recover the internal structure of jet has been tested by 
evaluating four jet shape variables that are discussed in the literature to be useful for analyzing the 
boosted objects or to perform jet tagging \cite{aakf:Altheimer, aaec:Gallicchio}. Here we briefly introduce these four jet shapes:


\begin{itemize}

 \item The jet mass which can also be used to identify the hadronic decays of boosted heavy particles \cite{aais:Abdesselam, aait:CMS}.

\item The $N$-subjettiness, \tauN, defined as~\cite{aakb:Thaler}
  %
  \begin{equation}
  \tauN = \frac{1}{d_{0}} \sum_k p_{\mathrm{T}k} \cdot  \text{min}(\Delta R_{1k}, \Delta R_{2k},...,\Delta R_{Nk})~\text{,~~with},
  ~~~d_{0}\equiv\sum_{k} p_{\mathrm{T}k}\cdot  R
  \label{eq:nsubj}
  \end{equation}
  %
  where $R$ is the distance parameter of the jet algorithm, $p_{\mathrm{T}k}$ is the transverse momentum 
of constituent~$k$ and $\Delta R_{ik}$ is the distance between a subjet~$i$ and a constituent~$k$. The $N$ subjets 
are defined by re-clustering the constituents of the jet with exclusive version of the 
\kt algorithm \cite{aaiy:Catani} and requiring that exactly $N$ subjets are found.
  Beside the $N$-subjettiness also the subjettiness ratio, $\tau_{MN} = \tau_M/\tau_N$, can be used to 
characterize the jet substructure. Typically, the three-to-two ratio, $\tauThrTwo=\tau_3/\tau_2$, is used 
which provides a good discrimination between standard QCD jets and jets formed e.g. by boosted top quarks. 
  
\item The \kt splitting scale, \DOneTwo, defined as~\cite{aakc:Butterworth}
  \begin{equation}
  \DOneTwo = \text{min}(\pt^1, \pt^2)\cdot \Delta R_{12},
  \label{eq:ktsplitting}
  \end{equation}
  \noindent
  where $\pt^1$ and $\pt^2$ are the transverse momenta of two subjets and $\Delta R_{12}$ is the distance 
between these two subjets. The two subjets are found by going back one step in the clustering history of 
the \kt algorithm. The variable \DOneTwo can be used to distinguish heavy-particle decays, which tend to be 
reasonably symmetric when the decay is to like-mass particles, from the largely asymmetric splittings that 
originate from QCD radiation in light-quark or gluon jets. 


\item The longitudinally invariant version of the planar flow, $\mathrm{Pf}$, defined as \cite{aaig:Soyez}:
\begin{equation}
\mathrm{Pf}=\frac{4\lambda_1\lambda_2}{(\lambda_1+\lambda_2)^2},
  \label{eq:planarFlow}
\end{equation}
where $\lambda_1$ and $\lambda_2$ are the eigenvalues of $2\times 2$ matrix:
\begin{equation}
M_{\alpha\beta}=\sum\limits_i \pti\cdot(\alpha_i-\alpha_{\mathrm{jet}})\cdot(\beta_i-\beta_{\mathrm{jet}}),
\end{equation}
  where $\alpha$ and $\beta$ correspond to rapidity or azimuth.


\end{itemize}


\subsection{Jet shape subtraction}
\label{sec:shapes}

The \subtraction method is tested on various combinations of signal samples, \pileup conditions, clustering algorithms, and jet shapes defined in Sec.~\ref{sec:definitions}. The details of the configuration are 
provided at the beginning of Sec.~\ref{sec:performance}. The \subtraction can recover the original jet shape with a good accuracy in all evaluated combinations.

  %
  \begin{figure}[!h]
  \centering
  \begin{subfigure}{0.328\textwidth}
    \includegraphics[width=\textwidth]{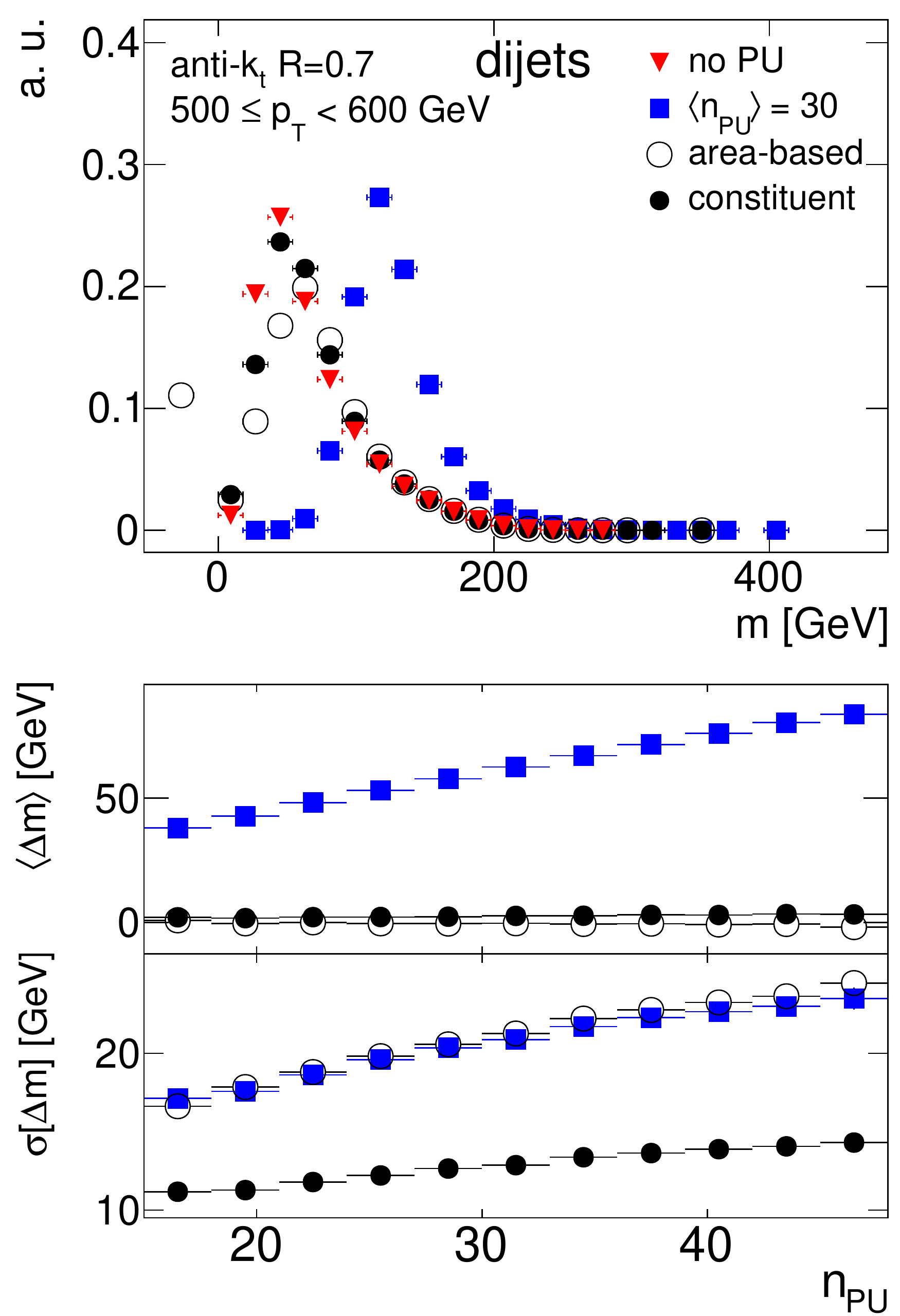}
    \label{fig:dijets:mass}
  \end{subfigure}
  \begin{subfigure}{0.328\textwidth}
    \includegraphics[width=\textwidth]{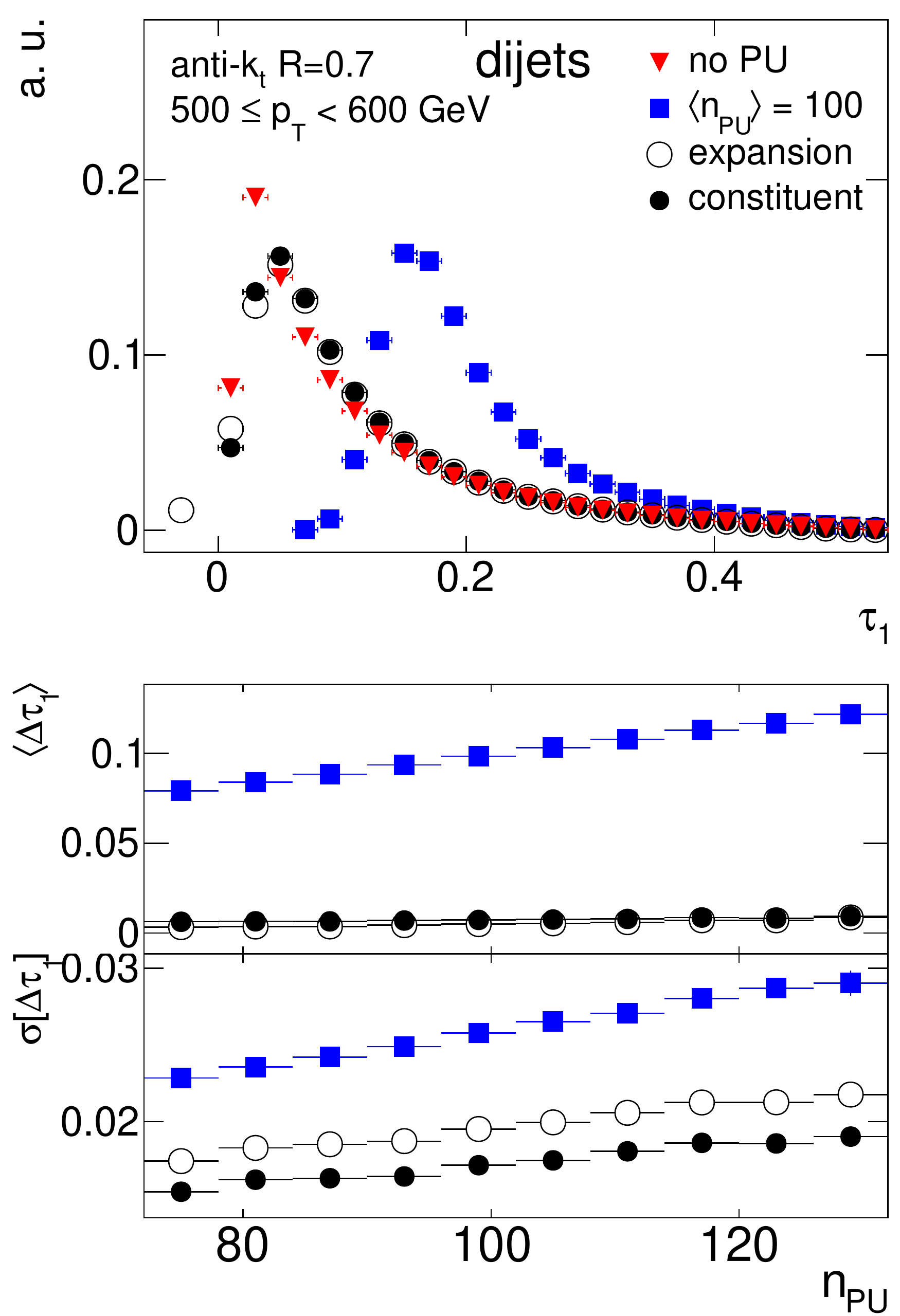}
    \label{fig:dijets:tau1}
  \end{subfigure}
  \begin{subfigure}{0.328\textwidth}
    \includegraphics[width=\textwidth]{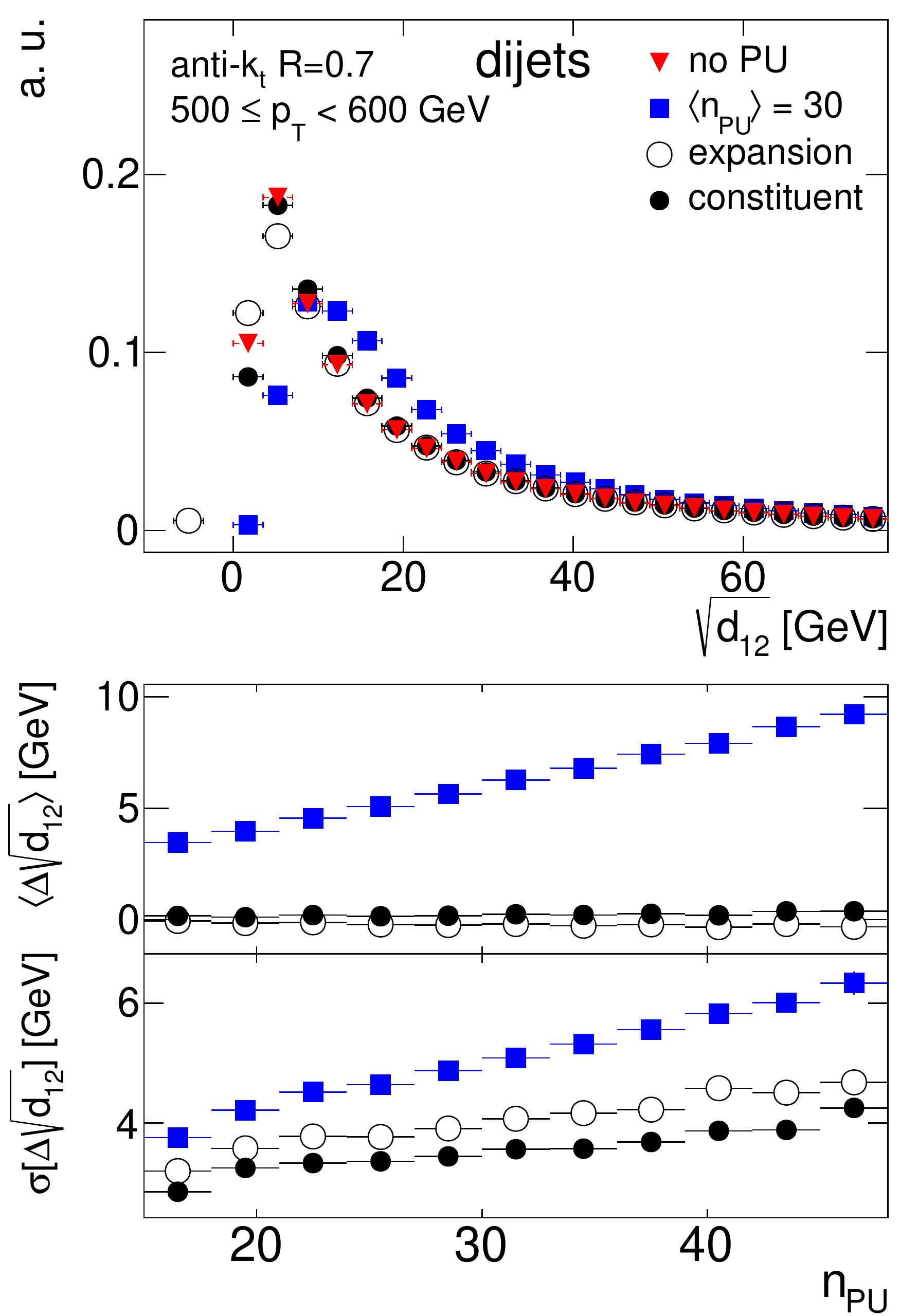}
    \label{fig:dijets:split12}
  \end{subfigure}
\\
  \centering
  \begin{subfigure}{0.328\textwidth}
    \includegraphics[width=\textwidth]{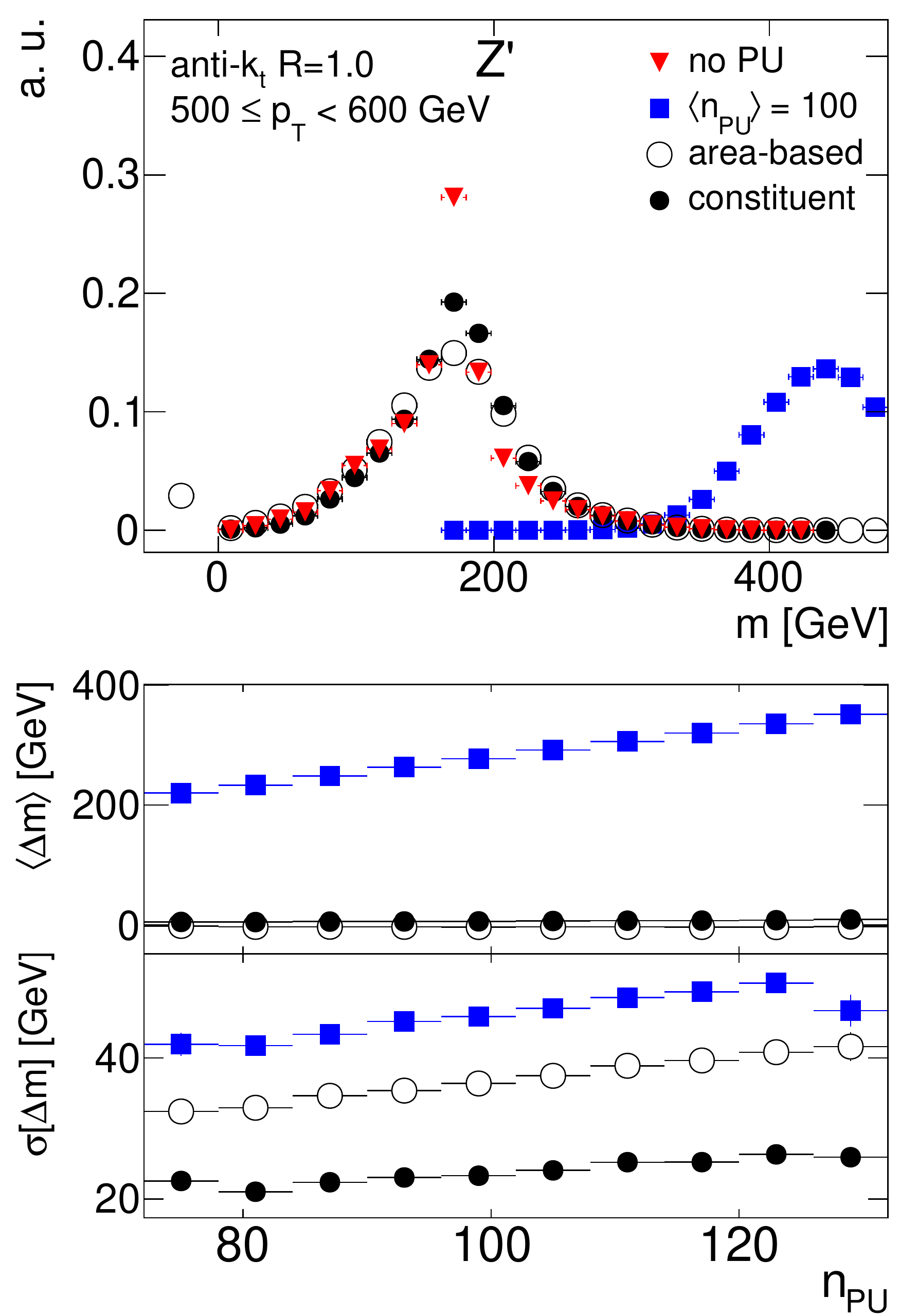}
    \label{fig:Zprime:mass}
  \end{subfigure}
  \begin{subfigure}{0.328\textwidth}
    \includegraphics[width=\textwidth]{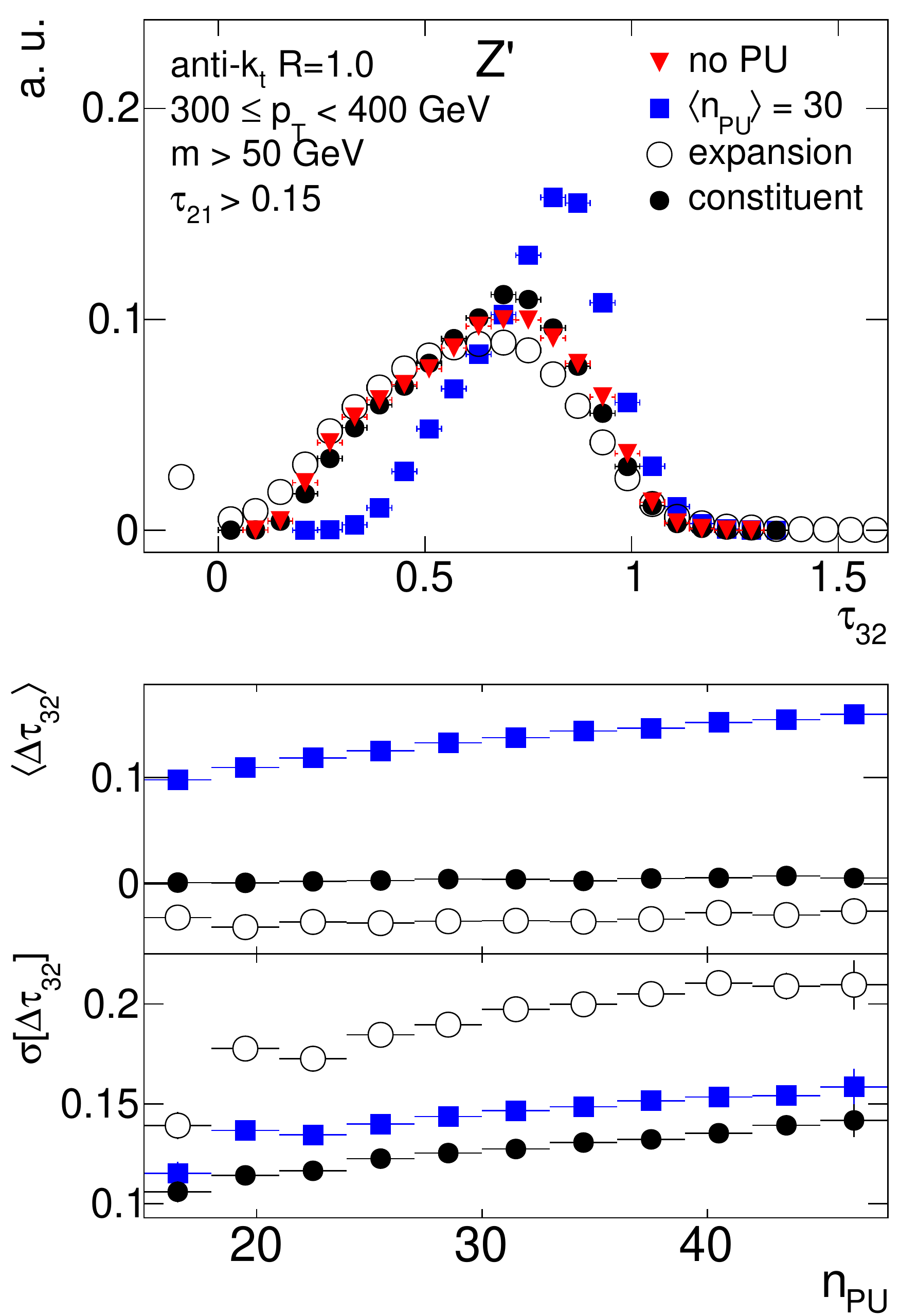}
    \label{fig:Zprime:tau3}
  \end{subfigure}
  \begin{subfigure}{0.328\textwidth}
    \includegraphics[width=\textwidth]{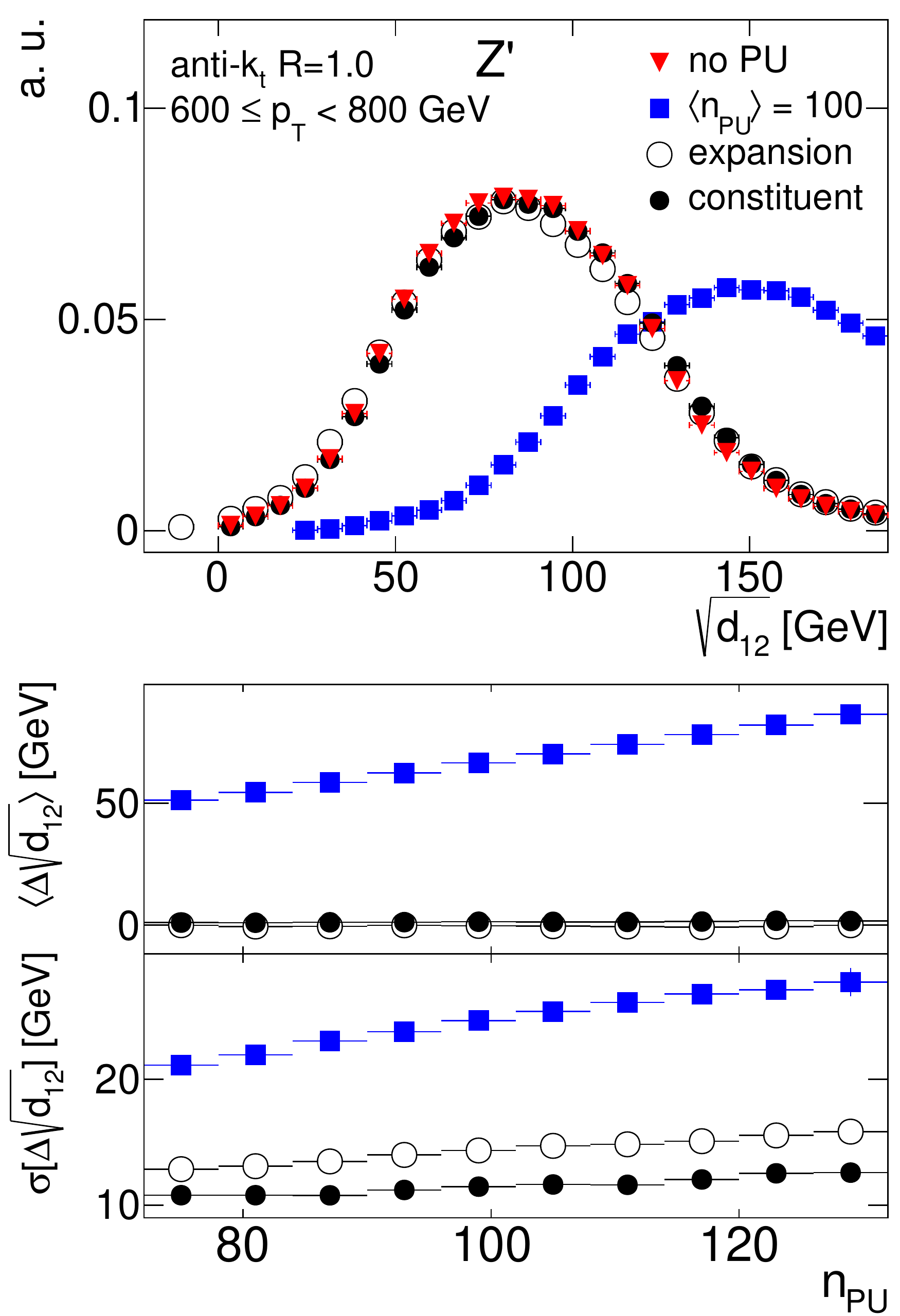}
    \label{fig:Zprime:split12}
  \end{subfigure}
  \caption{Performance of the constituent subtraction for jets clustered with the \akt algorithm. Red triangles show distribution without \pileup, blue squares show the uncorrected distribution with \pileup, open and closed circles show distributions corrected by the shape-expansion and \subtraction method, respectively. The \npu dependence of mean \meanDelta and standard deviation \sigmaDelta are shown in the lower panel for each jet shape.}  
  \label{fig:particles:antikt}
\end{figure}
%
%

A representative subset of performance plots for the \akt\ algorithm is shown in 
Fig.~\ref{fig:particles:antikt} for various jet \pt intervals\footnote{The \pt of the jets without \pileup is used to define the \pt intervals. The jets with \pileup and the corrected jets are matched to the original jets without \pileup. In some cases, additional cut is applied on mass or subjettiness ratio \tauTwoOne again on jets without \pileup.}. Four distributions of jet shapes are plotted for each presented plot: the original 
distribution (that is the distribution without \pileup), the distribution with \pileup, the 
distributions corrected by the \subtraction and the shape-expansion\footnote{For the jet mass, the area-based method using \equref{correctedFourMomentum} is used which is identical to the shape-expansion method.} methods. 
 To quantify precisely the performance of the correction, two quantities have been evaluated for the 
differences between jet shape $x$ and its original value without \pileup $x^{orig}$: the mean value of 
these differences $\meanDelta = \langle x - x^{orig} \rangle$ and the standard deviation of these 
differences $\sigmaDelta = \sigma[x - x^{orig}]$ which represents the resolution. For each combination of configurations, the uncorrected distributions differ significantly from the corresponding original distribution, and have a significant dependence on \npu. A substantial improvement is achieved by the \subtraction. The mean difference \meanDelta does not exhibit the \npu dependence and it is always centered near zero after the subtraction. The resolution \sigmaDelta is improved as well. The \subtraction method performs similarly or mostly better when compared to the shape-expansion method.
\begin{figure}[t]
  \centering
  \begin{subfigure}{0.328\textwidth}
    \includegraphics[width=\textwidth]{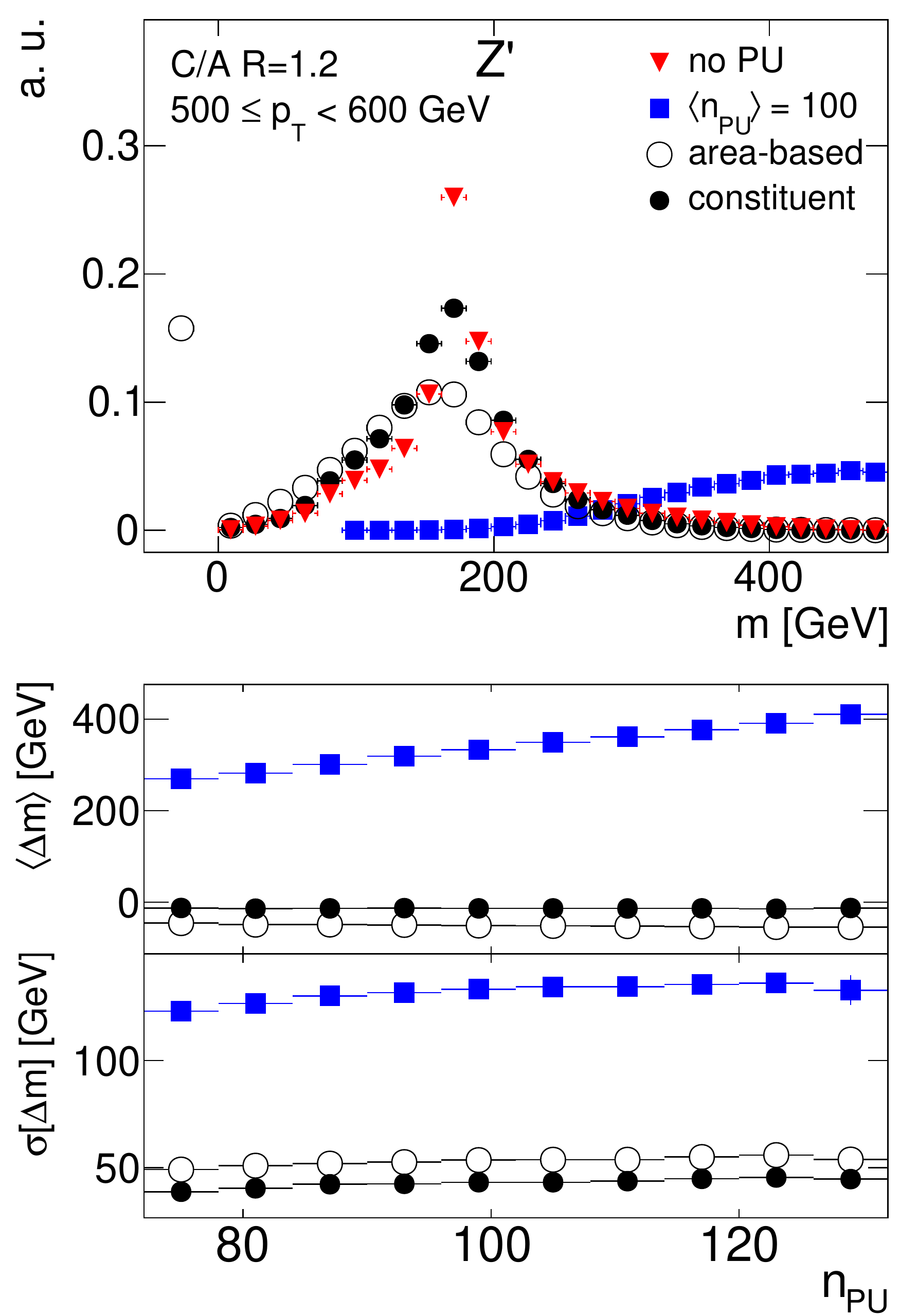}
    \label{fig:CA:Zprime:mass}
  \end{subfigure}
  \begin{subfigure}{0.328\textwidth}
    \includegraphics[width=\textwidth]{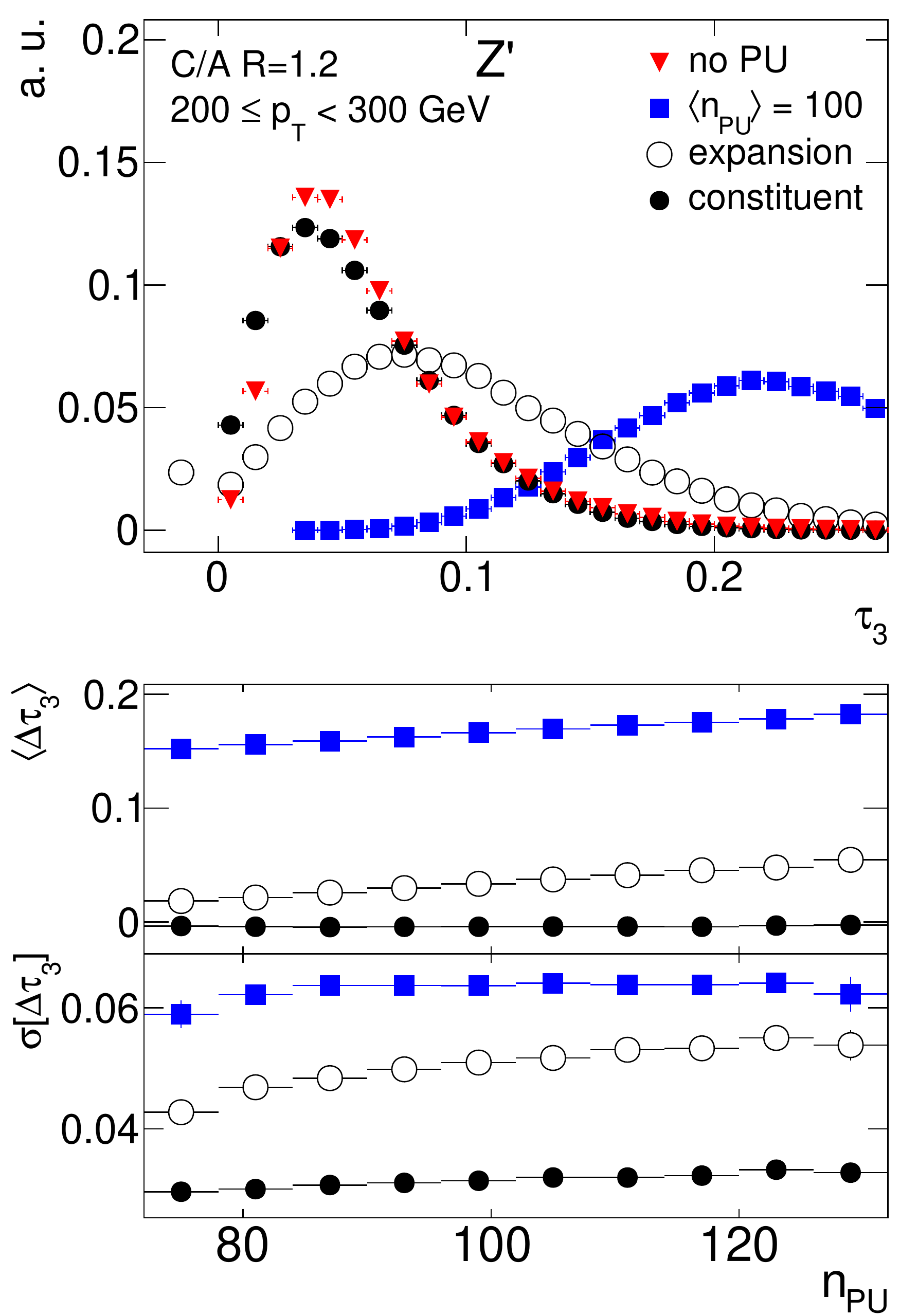}
    \label{fig:CA:Zprime:tau3}
  \end{subfigure}
  \begin{subfigure}{0.328\textwidth}
    \includegraphics[width=\textwidth]{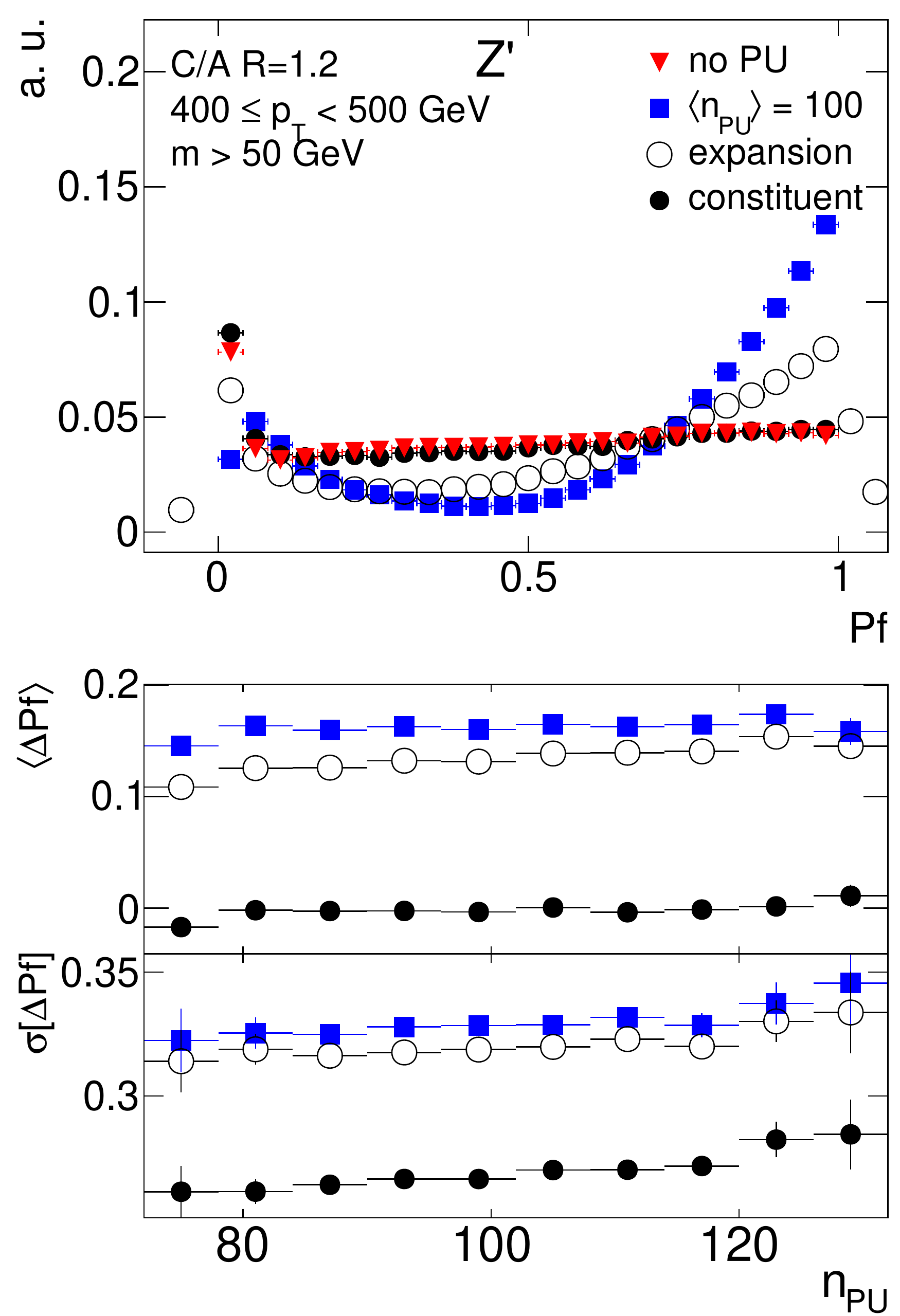}
    \label{fig:CA:Zprime:split12}
  \end{subfigure}
  \caption{Performance of the constituent subtraction for jets clustered with the \CamKt algorithm. Red triangles show distribution without \pileup, blue squares show the uncorrected distribution with \pileup, open and closed circles show distributions corrected by the shape-expansion and \subtraction method, respectively. The \npu dependence of mean \meanDelta and standard deviation \sigmaDelta are shown in the lower panel for each jet shape.}  
  \label{fig:particles:CA}
\end{figure}

\begin{figure}[t]
  \centering
  \begin{subfigure}{0.328\textwidth}
    \includegraphics[width=\textwidth]{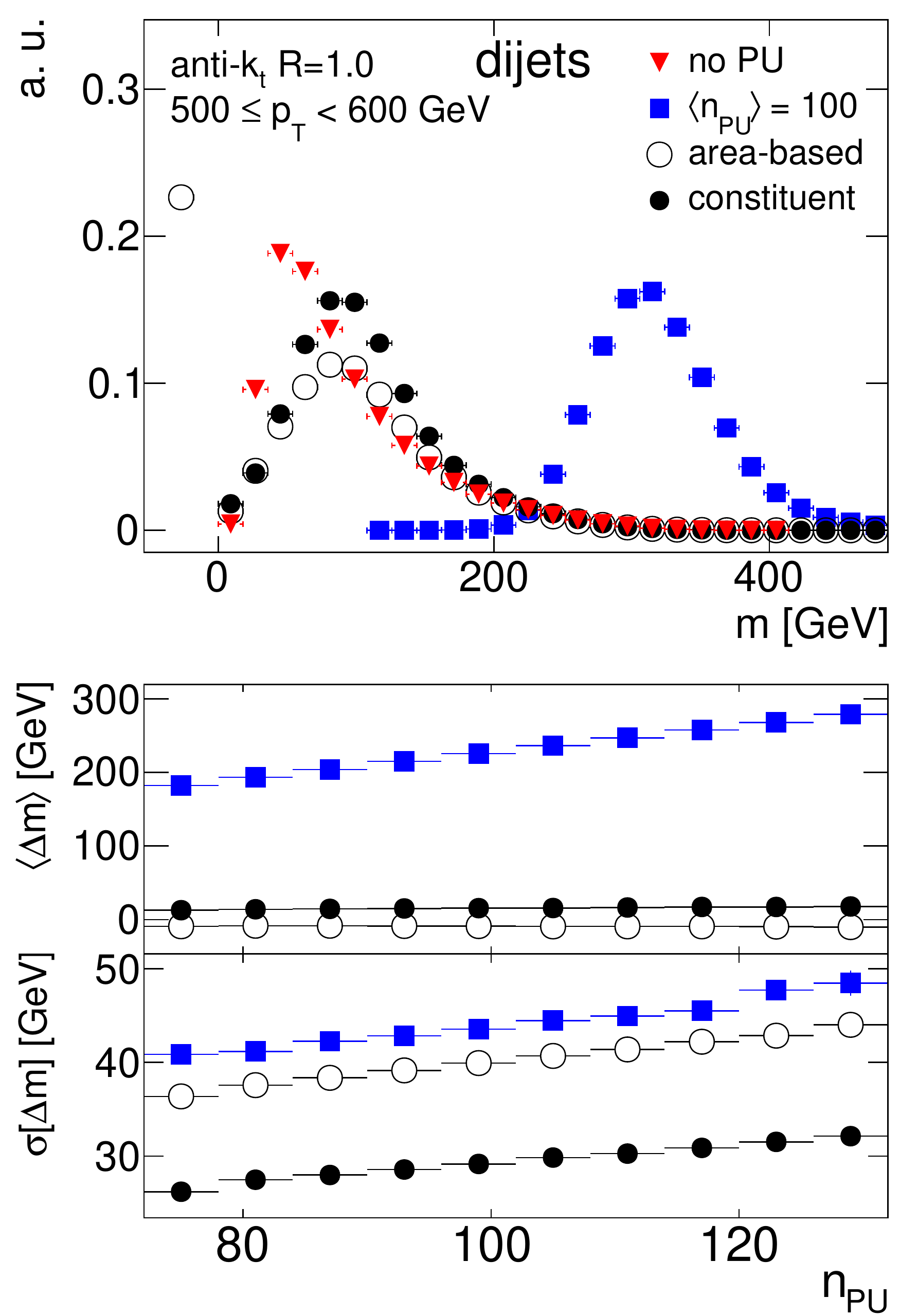}  
  \end{subfigure}
  \begin{subfigure}{0.328\textwidth}
    \includegraphics[width=\textwidth]{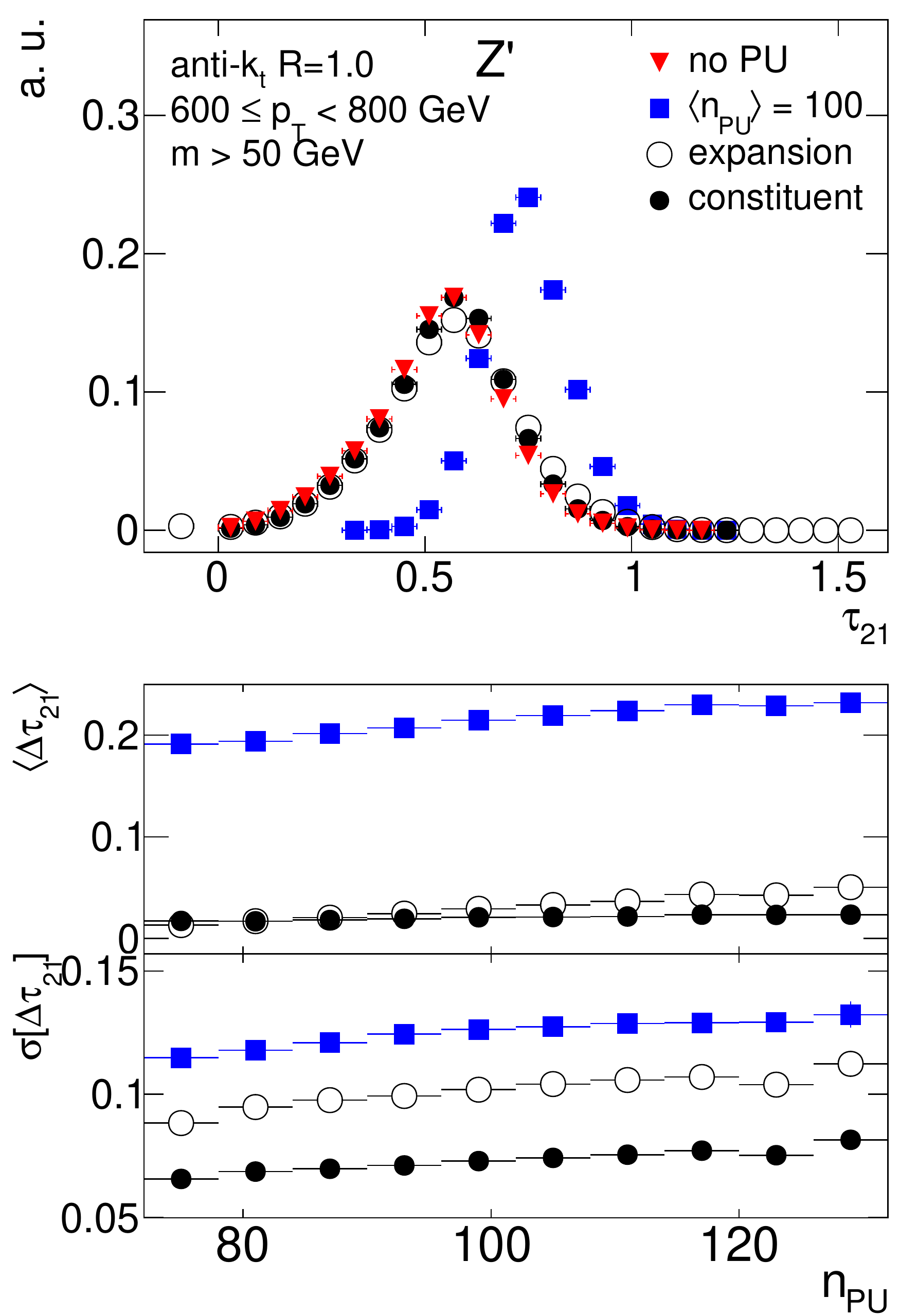}  
  \end{subfigure}
  \begin{subfigure}{0.328\textwidth}
    \includegraphics[width=\textwidth]{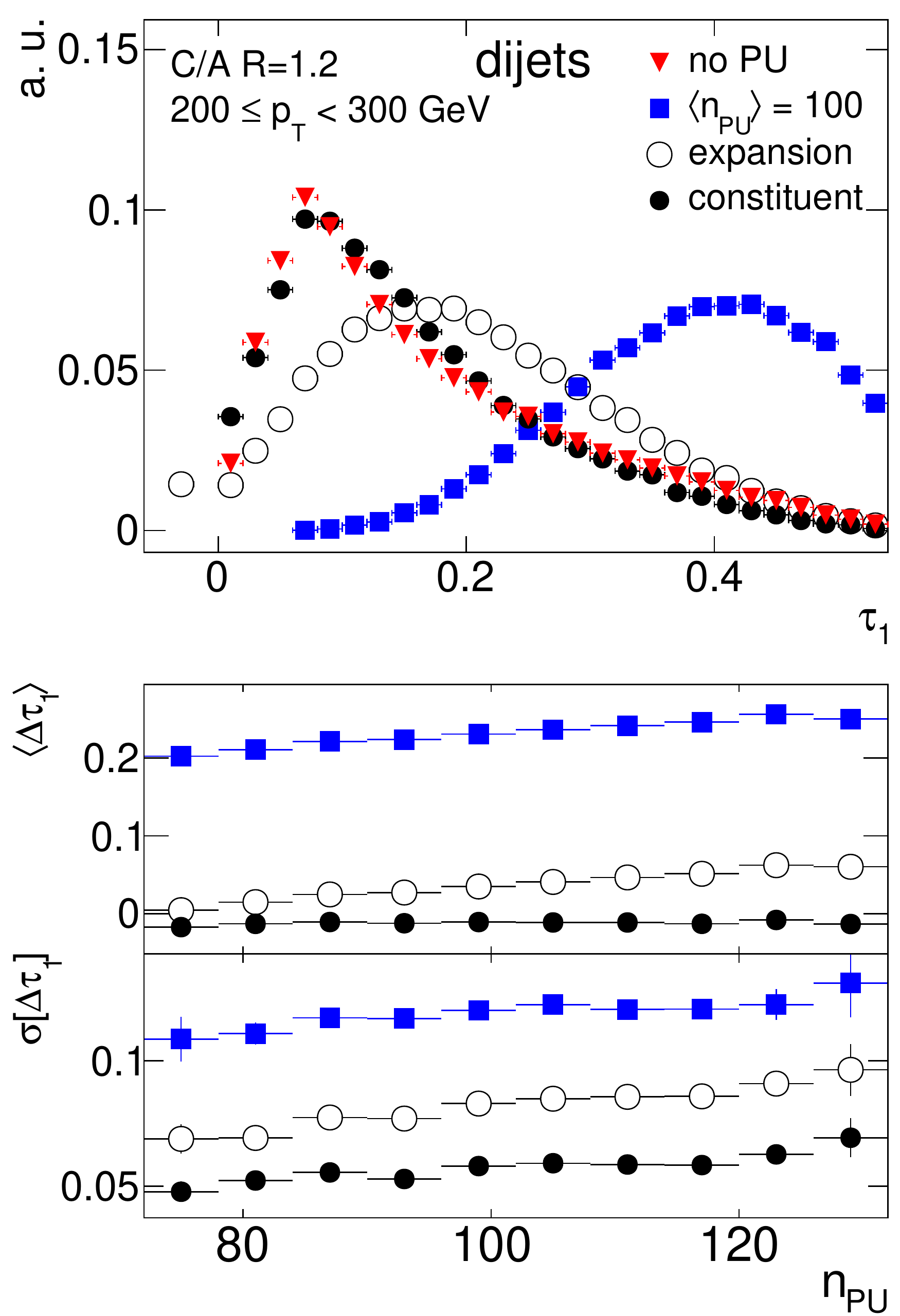}
  \end{subfigure}
  \caption{Performance of the \subtraction in events simulating a segmented detector. Red triangles show distribution without \pileup, blue squares show the uncorrected distribution with \pileup, open and closed circles show distributions corrected by the shape-expansion and \subtraction method, respectively. The \npu dependence of mean \meanDelta and standard deviation \sigmaDelta are shown in the lower panel for each jet shape.}  
  \label{fig:griddedParticles}
\end{figure}

For any of the studied jet shapes, the shape-expansion method can lead to negative corrected jet shapes 
that are unphysical. To better visualize the contribution of such values, the first bin with negative jet 
shape in plots of Figs.~\ref{fig:particles:antikt}-\ref{fig:griddedParticles} is set to the fraction of negatively corrected jet 
shapes\footnote{For the shape-expansion correction of ratios \tauTwoOne or \tauThrTwo, the 
numerator and denominator are corrected individually. When at least one of these corrected variables is negative, 
the corrected \tauTwoOne or \tauThrTwo is counted as negative. For the calculation of the mean and 
resolution of \tauTwoOne or \tauThrTwo, the negative values are not used. For any other jet shapes, the 
negative values are set to zero so that they do not bias the mean and resolution.}. Unphysical values can 
also occur in the case of the area-based correction of the jet mass when the corrected energy is smaller than 
the magnitude of corrected momentum. Again, the negative bin represents the fraction of such jets. The 
fraction of unphysical jet shapes obtained from the shape-expansion method reaches up to $\sim12\%$ depending on 
the \pt interval and the type of the jet shape.

The \subtraction method has been tested also on the jets clustered with \CamKt algorithm which is often employed in various studies 
of the jet substructure and boosted objects \cite{aakf:Altheimer}.  The clustering in the \CamKt algorithm is 
based purely on the geometry and thus it leads to jet with a different jet area compared to the \akt 
algorithm \cite{aabb:Cacciari}. The performance of the \subtraction method for \CamKt algorithm with distance 
parameter $R=1.2$ is shown in Fig.~\ref{fig:particles:CA}. 
  For this configuration, the impact of the \pileup on jet shapes is much stronger compared to the 
configuration with the \akt algorithm.
  The \subtraction can recover the original distributions and it exhibits significantly better ability to 
subtract the \pileup compared to the shape-expansion method.

Further, the \subtraction method has been tested on jets reconstructed in events run through a simple simulation of 
a segmented detector. In this simulation, the $\eta - \phi$ plane is divided into cells of size 
$0.1\times0.1$. Particles pointing to the same cell are combined into one new effective particle by summing 
their energies. The mass of the cell is set to zero and its $\eta - \phi$ position is set to the center 
of the cell. These new effective particles have the same properties as the calorimeter clusters or towers used 
in real experiments and they are a combination of the \pileup and signal. The jet finding algorithm runs over 
these events delivering jets that are corrected in the same way as for events composed of standard 
particles. A typical example of the performance of the subtraction methods in case of this simulation is shown in Fig. 
\ref{fig:griddedParticles}. 
The \subtraction exhibits very similar performance as without the detector simulation which also applies 
 for the shape-expansion method while the \subtraction again outperforms the shape-expansion method.
%

\begin{figure}[t]
  \centering
 \begin{subfigure}{0.48\textwidth}
    \includegraphics[width=\textwidth]{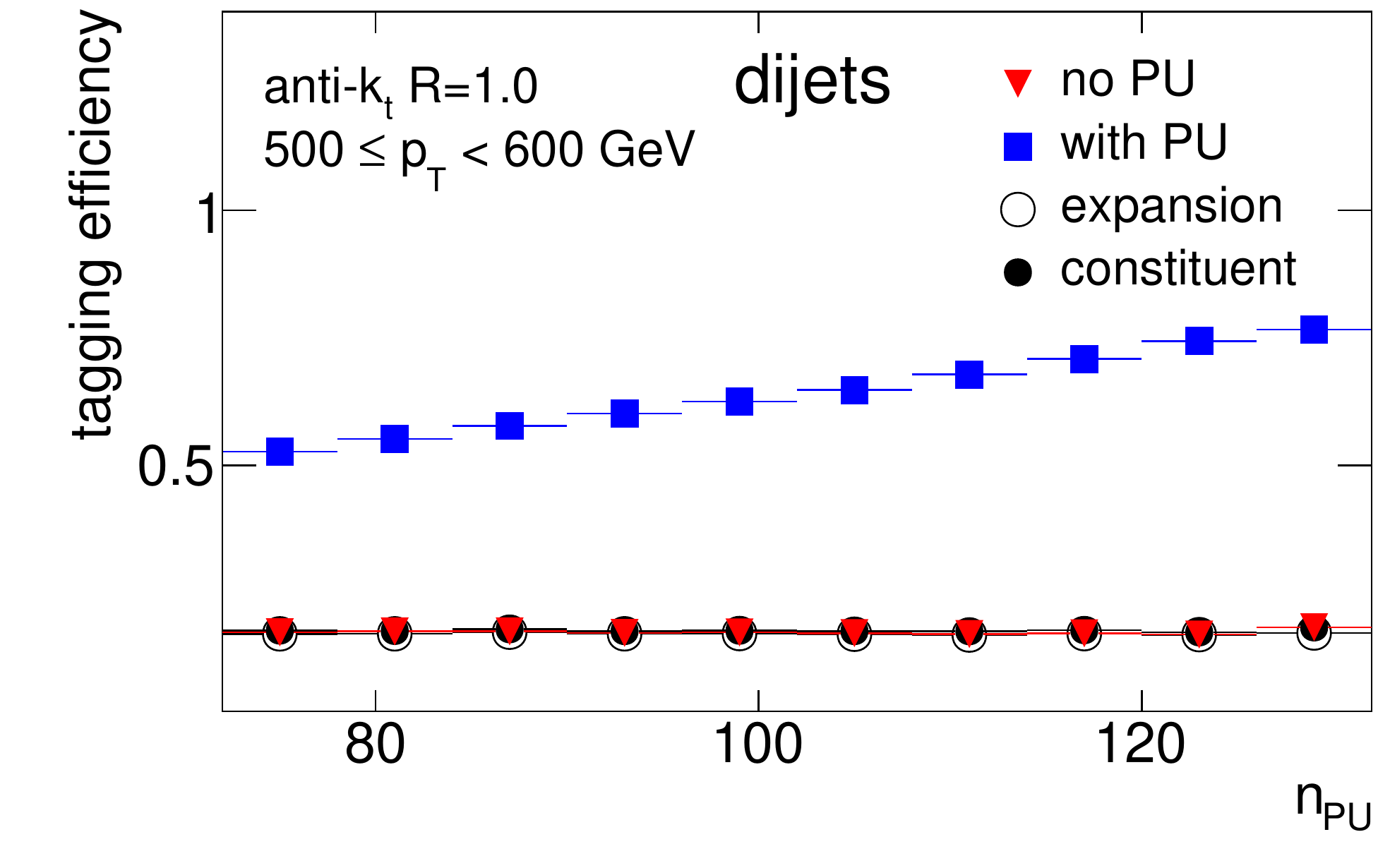}
  \end{subfigure}
  \begin{subfigure}{0.48\textwidth}
    \includegraphics[width=\textwidth]{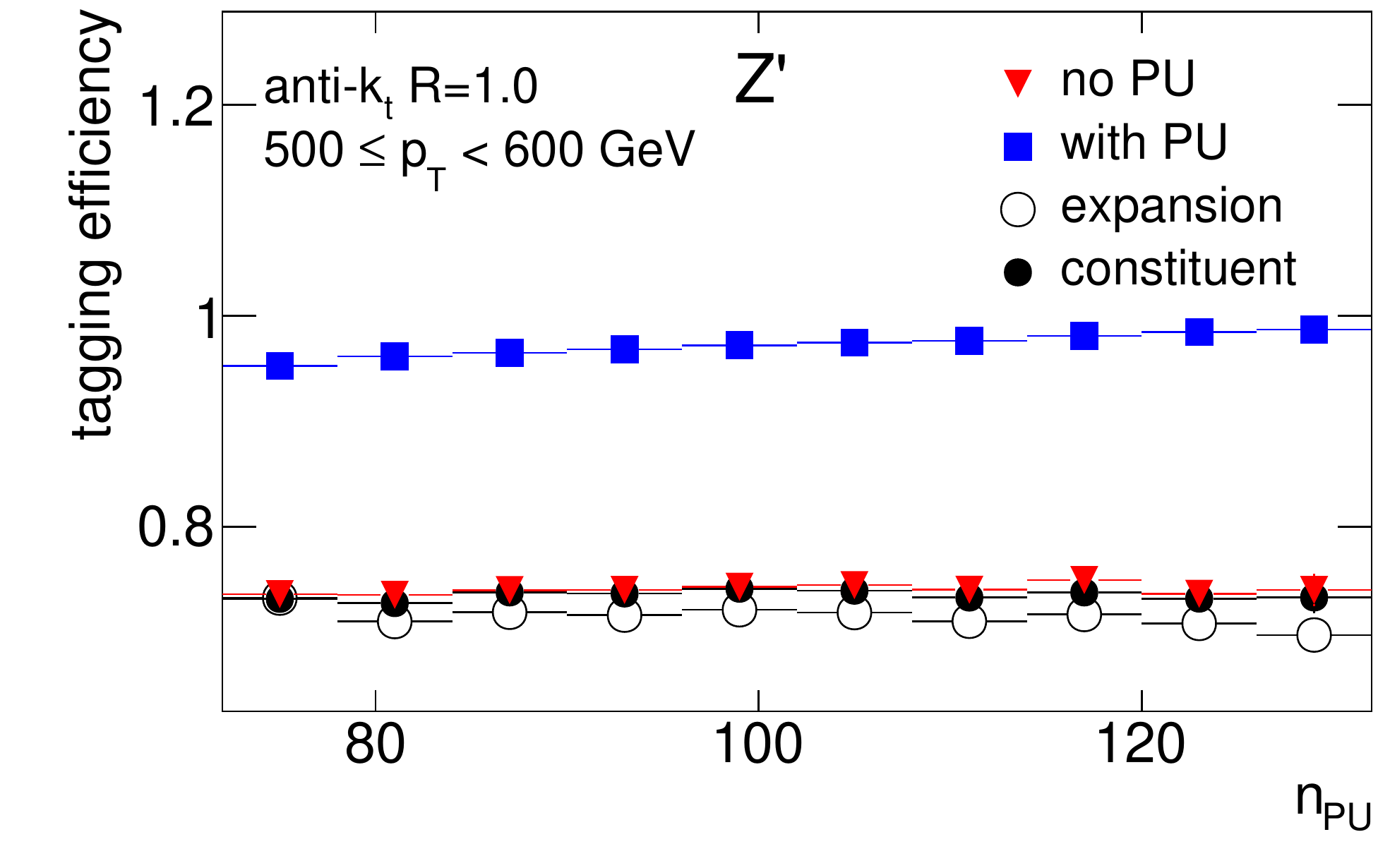}
  \end{subfigure}
  \caption{Tagging efficiencies for background (left) and signal (right) samples using tagging cut $\DOneTwo>50 \GeV$ in events simulating a segmented detector. Red triangles show the tagging efficiency for jets without \pileup. Blue squares show the tagging efficiency for jets with \pileup, open and closed circles show tagging efficiency for jets corrected by the shape-expansion method and \subtraction method, respectively.}  
  \label{fig:efficiencies}
\end{figure}

An important test is to evaluate the jet tagging performance. The splitting scale, \DOneTwo, is used to 
tag the boosted top quarks from the $Z'$ decay and tagging efficiencies have been evaluated. The signal 
sample can be compared with the di-jet sample which in this case provides a reasonable estimate of the 
background for the $Z' \rightarrow t\bar{t}$ decay. The result is shown on Fig. \ref{fig:efficiencies}. One can 
see that both the \subtraction and shape-expansion can achieve the same tagging efficiency as in the case 
of no \pileup.

        %
%

The above presented results demonstrate the stability and good performance of the \subtraction method.


\section{Conclusions}
\label{sec:conclusions}

We have introduced a new tool to correct for the \pileup in high-luminosity LHC running that represents an extension and a simplification of the current state of the art. The \subtraction method operates at the level of the jet constituents and provides both a performance improvement and a simplification compared to existing methods: the precision of the reconstruction of jet shapes is improved as well as the speed of the correction itself. 

The \subtraction method is tested by evaluating the \pileup dependence, and other key metrics, of several jet shapes and jet kinematics using multiple jet definitions. Improvements are demonstrated in both the reduction of fluctuations in the resulting jet shapes as a function of \pileup and the ability to remove the \pileup dependence of the corrected quantities. It also provides a better jet position resolution and jet finding efficiency for reconstructing jets from boosted objects, which directly impacts the experimental sensitivity to new physics. 

Since the correction proceeds without knowledge of a jet algorithm, a novel application of this approach would be to correct the whole event prior to jet finding. Not only would this have implications for the technical performance and computational resources currently devoted to evaluation the \pileup subtraction for each jet individually, but it also has the potential to improve the determination of the missing transverse energy. Furthermore, the \subtraction approach can be used to correct for the underlying event in heavy ion collisions.

It is very important that these studies be verified by the experiments using both fully simulated data samples of signal and background events, as well as \textit{in situ} studies using data. Given the excellent correspondence between the similar preliminary studies of earlier \pileup correction methods and the experimental reality, we expect that the performance observed here is quantitatively representative of what can be achieved by the experimental collaborations. Nonetheless, any new approach must be vetted and tested thoroughly.

\acknowledgments

The authors would like to thank Gregory Soyez and Gavin Salam for their feedback and assistance in understanding and using the shape-expansion tools. The work is supported by Charles University in Prague, projects PRVOUK P45, UNCE 204020/2012, and GA UK No 320813. DWM is supported in part by the Neubauer Family Foundation Fellows Program for Assistant Professors.


\end{document}